# Synthetic Speaking Children – Why We Need Them and How to Make Them

Muhammad Ali Farooq, Dan Bigioi, Rishabh Jain, Wang Yao, Mariam Yiwere, Peter Corcoran
*College of Science and Engineering, University of Galway,* Galway, Ireland
muhammadali.farooq@universityofgalway.ie, d.bigioi1@nuigalway.ie, rishabh.jain@universityofgalway.ie,
w.yao2@universityofgalway.ie, mariam.yiwere@universityofgalway.ie, peter.corcoran@nuigalway.ie

*Abstract*—Contemporary Human-Computer Interaction (HCI) research relies primarily on neural network models for machine vision and speech understanding of a system user. Such models require extensively annotated training datasets for optimal performance and when building interfaces for users from a vulnerable population such as young children, GDPR introduces significant complexities in data collection, management, and processing. Motivated by the training needs of an Edge-AI smart-toy platform this research explores the latest advances in generative neural technologies and provides a working proof-of-concept of a controllable data-generation pipeline for speech-driven facial training data at scale. In this context, we demonstrate how StyleGAN-2 can be fine-tuned to create a gender-balanced dataset of children's faces. This dataset includes a variety of controllable factors such as facial expressions, age variations, facial poses, and even speech-driven animations with realistic lip synchronization. By combining generative text-to-speech models for child voice synthesis and a 3D landmark-based talking heads pipeline, we can generate highly realistic, entirely synthetic, talking child video clips. These video clips can provide valuable, and controllable, synthetic training data for neural network models, bridging the gap when real data is scarce or restricted due to privacy regulations.

## I. Introduction

In the dynamic landscape of human-centric machine vision and speech analysis, researchers are frequently confronted with substantial challenges stemming from GDPR guidelines. Contemporary research heavily relies on neural network models and the availability of extensive training datasets to attain optimal performance. However, when the research focus shifts to the development of Human-Computer Interaction (HCI) interfaces and necessitates data from vulnerable populations, particularly young children, to train Edge-AI HCI models, GDPR introduces a myriad of complexities associated with data collection, management, and processing. These complexities are particularly pronounced when dealing with real data involving children, where stringent privacy regulations come into play.

In response to these challenges, recent advancements in Generative Adversarial Networks (GANs) and other generative neural technologies have emerged as promising solutions for generating data at scale. In this context, this paper introduces an innovative approach that leverages the power of such technologies. Motivated by the need for training data for an Edge-AI-based smart-toy platform [1] we demonstrate the adaptability of StyleGAN-2 [2], [3], a state-of-the-art generative neural architecture, to craft a gender-balanced dataset of synthetic children's faces. This dataset offers nuanced control over various critical attributes, including facial expressions, age variations, [1] facial poses, and the synchronization of facial movements with speech-driven animations, culminating in a collection of strikingly realistic videos.

Going beyond visual representation, our exploration extends into the domain of voice generation. By incorporating techniques such as FastPitch, advanced voice augmentation, and generative text-to-speech models, we achieve the ability to synthesize authentic children's voices, replete with their distinctive qualities. These voices, when seamlessly integrated with our StyleGAN-2 framework and speech-driven neural lip synchronization models, empower us to create highly realistic, entirely synthetic talking child videos.

These synthetic videos represent a pragmatic solution to data scarcity or stringent privacy regulations like GDPR. In addition to their applications in research and development, these videos serve as valuable training data for neural network models in a practical use case, such as an Edge-AI smart-toy platform. In developing these tools our focus has been on scaling to enable controllable data generation at scale. Thus spoken phrases can be employed in combination with a set of synthetic voices and multiple seed faces to fine-tune the computer vision and automated speech recognition models that operate on the smart-toy platform. This is useful, for example, to test how well the smart toy can detect the emotional state of a child or respond to variations in the command set for an interactive play activity. Gathering such data from children and directing their responses in a controlled laboratory environment is both time-consuming and costly.

The rest of this paper is devoted to providing a detailed explanation of our synesthetic child media generation pipeline starting with Section II which covers the creation of synthetic face samples, Section III which covers the generation of synthetic voice samples with FastSpeech 2, and Section IV which details how videos are created using MakeItTalk. Section V offers details on our experimental setup, with Section VI concluding the contents of this work.

## II. An Overview of ChildGAN

The first step includes generating large-scale synthetic child facial data using advanced data augmentation methods. This is achieved by fine-tuning StyleGAN2 [2] for generating photo-realistic child data samples. This new synthetic child dataset is referred to as ChildGAN [5].

### A. Training Methodology

StyleGAN2 is fine-tuned by using a transfer learning-based methodology. Transfer learning on GANs is a powerful technique, especially when there are limited amounts of data and computational resources. It allows us to leverage the knowledge and representations learned by pre-trained CNN models and adapt them for new tasks and domains. For this study, we have trained the adapted StyleGAN2 model using a synthetic child dataset as the seed data. The original seed data is taken from adult data samples and is transformed using various algorithms, including GANs and Android-based mobile apps that create a child facial image from an input adult image. The overall process involves the fine-tuning of the StyleGAN2 generator, and discriminator, by using an adversarial training methodology. The complete training method is detailed in the work of [5].

### B. Dataset and Quality Checks

We assess the quality of the synthetic data by employing various computer vision methods, ensuring the excellence and detail of the facial features in the generated data. To accomplish this, we utilize a combination of qualitative and quantitative metrics. These metrics encompass essential tests, such as face localization and facial landmark detection using the DLIB framework. Another crucial assessment involves the computation of the cosine similarity index, facilitated by ArcFace [6], to measure the similarity in identity among synthetic child faces. Additionally, we validate the quality of the artificial child faces for downstream applications by conducting tests with child gender classifiers, allowing us to evaluate the performance of these classifiers on real data [5]. In Fig. 1, we present a visual representation illustrating synthetic facial data featuring both boys and girls, generated through the ChildGAN network.

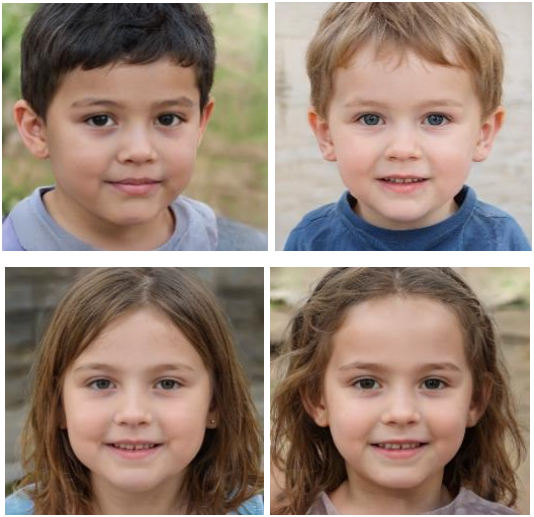

Fig. 1. Four distinct child facial samples of boys and girls generated from ChildGAN by finetuning StyleGAN2 with latent space editing.

### C. Facial Transformations and Tools

The rendered synthetic data is further transformed to incorporate various smart transformations that can be used for diversified real-world computer vision applications. This is achieved by using the latent space editing feature in StyleGAN2. These transformations include eye blinking effects, age progression, directional lighting conditions covering different facial angles, facial expressions, head pose variations, and lastly hair and skin tone digitization. The complete dataset along with pretrained models are open sourced which can be used for more extensive data generation, further experimental analysis, and other related downstream tasks. Fig. 2 shows two different smart data transformations done via latent space editing [8] and relighting [7] based deep learning networks.

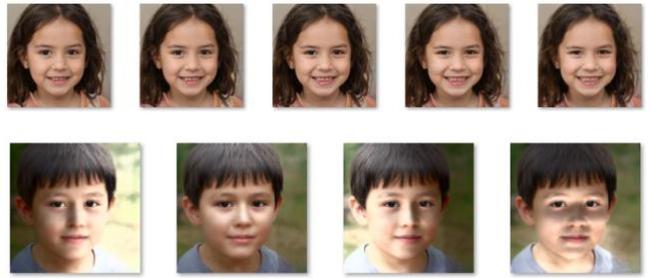

Fig. 2. Rendered child facial data with further advanced data augmentation results. The first row shows neutral to happy facial transformation on the generated girl subject and the second row depicts four directional lighting conditions embedded via [6] on the sample boy subject.

## III. SYNTHESIZING CHILDREN'S VOICES

Employing Cleese-based pitch augmentation [9], FastPitch TTS [10], and Tacotron2 TTS [11] for generating synthetic child speech holds immense potential for research applications. Cleese-based pitch augmentation allows for precise control over pitch contours, enabling the creation of highly realistic child voices with varying age and gender characteristics. FastPitch and Tacotron2, as cutting-edge TTS models, ensure the conversion of text into natural and expressive child-like speech. By leveraging synthetic data, researchers can conduct experiments without ethical concerns associated with using real child participants, while also ensuring reproducibility and standardization.

### A. Text-to-Speech Child Voice Synthesis using Tacotron 2

The multi-speaker TTS model [12] consists of three distinct neural network models, each addressing a specific subtask: the Speaker Encoder for speaker verification, the Acoustic model for spectrogram synthesis, and the WaveRNN Vocoder for audio waveform generation.

The Speaker Encoder is trained using a combination of adult and child speech data from various datasets. It utilizes the Generalized End-to-End (GE2E) [13] loss to generate fixed-dimensional speaker embeddings. These embeddings enable the model to effectively distinguish between different speakers, allowing for better generalization across various voices. During training, complete utterances are segmented into partial utterances of 1.6 seconds, and the encoder is optimized over GE2E loss to ensure similar voices are mapped closer together in a latent space representation.

The Tacotron 2 Acoustic model [11], originally designed for single-speaker TTS, is adapted for multi-speaker functionality by incorporating the speaker embeddings alongside the text embeddings. The model is first trained with adult speech data and then fine-tuned with child speech data. The combination of speaker and text embeddings enhances the model's capability to generate spectrograms from input text conditioned on the specific speaker identity.

For audio waveform generation, the researchers employ the WaveRNN Vocoder [14], an improvement over the WaveNet model. WaveRNN is particularly chosen for its ability to perform sequential modeling of audio from mel-spectrograms. It utilizes a Gated Recurrent Unit (GRU) to replace convolutions used in WaveNet [15], reducing sampling time while maintaining high output quality. The vocoder is trained on adult speech data and proves to be effective even with unseen speakers in multi-speaker models.

The proposed approach exhibited promising results in generating high-quality synthetic child voices which was verified using various subjective and objective evaluations.

*B. Augmentation Techniques for Adult Voices*

To generate synthetic child-like speech data from existing adult speech, a python-based sound manipulation toolkit known as Combinatorial Expressive Speech Engine (CLEESE) [9] is used to augment the adult speech data to make them closer to child voices. A d-vector based speaker encoder is used to compare the adult speaker embeddings to the mean child embedding to select the adult speakers most similar/proximate child speakers for the augmentation based on the cosine similarity metric. Specifically, the pitch and speaking rate of the selected adult speakers are raised and slowed down through the CLEESE pitch-shift and time-stretch transformations respectively, causing them to sound more child-like.

Objective and subjective (Mean Opinion Score (MOS)) evaluations performed showed that the Cleese-based augmentation approach successfully tuned the adult voices to sound child-like; however, due to the adult linguistic content and the absence of child-like prosodic features such as long pauses and "stammering", the augmented speech lacked the naturalness of real child speech. The evaluations also revealed that adult female speakers generally provided a better starting point for the augmentations as compared to adult male speakers. Overall, the average MOS score of 3.7 was reported for how convincing the augmented speech samples are as child speech and 4.6 for how intelligible the augmented speech is, for the best set of augmentation parameters. The work has been submitted to the IEEE ACCESS journal and is currently in the review process. In future work, we plan to improve the time-stretch transformation (speaking rate augmentation) in addition to modeling the child-like prosody and other paralinguistic features as part of the current augmentation approach; this is expected to improve the similarity between augmented speech and real child speech.

*C. Using FastPitch to Synthesize Child Voices*

A transfer learning pipeline is used for generating synthetic child voices using the Fastpitch TTS [10] model. The process involves pretraining the model with the LibriTTS [16] dataset, which includes diverse adult speech data. Then, the model is fine-tuned on a small subset of child speech data (MyST dataset [17]) to capture the acoustic properties and pitch contours specific to child speech. The finetuning pipeline is consistent with previous approaches using Tacotron 2 [11]. The vocoder used for generating high-quality speech waveforms is WaveGlow, which operates based on a generative flow-based model architecture. The WaveGlow [18] model is trained on the LibriTTS adult speech data and is employed as a universal vocoder for synthetic child voices.

Objective evaluations on the naturalness and intelligibility of the generated speech are conducted, comparing the Fastpitch model's performance with Tacotron 2 [11] for child speech synthesis. Moreover, speaker similarity verification using a pretrained speaker verification system shows that the synthetically generated child speech is close to real speech in terms of speaker similarity. This methodology successfully synthesizes realistic child voices, and the experimental results support the effectiveness of the Fastpitch model in generating high-quality synthetic child speech.

IV. SYNTHETIC TALKING HEAD GENERATION

Talking head generation presents a multitude of intricate challenges, including the precise synchronization of lip movements with speech, the maintenance of natural facial expressions throughout the animation process, and the overall cohesiveness of facial dynamics with the spoken content. Additionally, issues related to data quality, articulatory variation across different speakers, and achieving a high degree of realism in the generated faces are all formidable hurdles in this domain. These hurdles are further amplified when trying to generate synthetic child data, as children's speech and facial expressions exhibit unique characteristics and idiosyncrasies that demand specialized handling. Children's facial features and articulatory patterns differ significantly from those of adults, making it essential to tailor the synthesis process to capture these nuances accurately. Ensuring the generated child faces are both age-appropriate and realistic adds an extra layer of complexity to the task.

Existing talking head generation approaches often overlook these specific challenges, primarily because they are predominantly trained on adult data. Consequently, adapting such models for child-focused applications presents an open challenge in the field. The need to address the distinct nuances of child speech and facial expressions, while also maintaining the integrity of age-appropriate visual representations, underscores the gap that our research aims to highlight. With that in mind, it must be made clear that while our research does not explicitly tackle the unique issues associated with child-specific talking head generation, it serves as a foundational step toward exploring the potential of synthesizing child faces in this context.

In the current work, our focus is primarily on leveraging established techniques to create a synthetic dataset that can potentially catalyze further investigations and innovations in the field, paving the way for more specialized solutions to address the intricacies of child-focused talking head generation.

*A. Rendering the Synthetic Talking Child Faces*

MakeItTalk [4] is a structural-based talking head generation approach that works by generating a sequence of sparse 3D facial landmarks given a driving audio signal as input, followed by an image-to-image translation-based rendering step that generates realistic video frames from the landmark sequence and an input "seed" image. Two models are used to accomplish this, an audio-aware LSTM-based model for generating landmark sequences time-aligned to the input audio signal, and a Pix2Pix-based image rendering network.

In theory, any recent talking head generation model can be used for this step, however, we chose to use MakeItTalk [4] as it displayed remarkable robustness when exposed to synthetic voice data as input. We theorize that this can be attributed to how MakeItTalk handles audio input. Specifically, it adopts a process that disentangles the input speech into two distinct latent representations: a content embedding and a speaker ID embedding. This appears to bolster the model's adaptability and its ability to generate accurate and contextually relevant facial landmarks, making it an ideal choice for our use case.

In total, we generate and provide 20 synthetic child-speaking videos comprising of both boy's and girl's facial samples, each uniquely characterized by identities meticulously crafted through the ChildGAN model, and speech samples synthesized in accordance with the detailed process outlined in Section III. A high-level overview of this pipeline is depicted in Fig. 3.

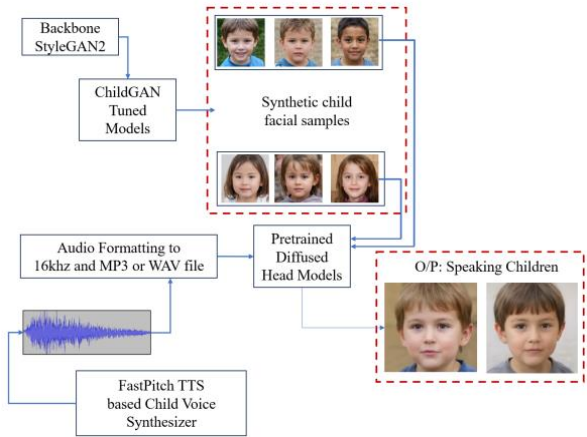

Fig. 3. Block diagram representing the pipeline adapted for generating 3D synthetic child speaking clips.

These synthetic videos display the capabilities of our synthesis pipeline. By offering this framework, we aim to facilitate a deeper understanding of child-like facial expressions and their correlation with speech, while also providing a valuable resource for researchers across various fields. Furthermore, we make all code, and scripts associated with this research publicly available.

*B. Evaluating Synthetic Videos*

We have abstained from presenting a formal evaluation of the synthetic videos we provide in this study. The rationale behind this decision is rooted in the fact that the videos we generate using our pipeline share a similar nature with those produced by MakeItTalk, which serves as a fundamental component of our methodology. Since MakeItTalk is a well-established framework for talking head generation with a recognized set of evaluation metrics and benchmarks, it offers a reliable reference point for the assessment of synthetic videos generated through our approach.

For a comprehensive and in-depth analysis of the specific characteristics, quality, and performance of the videos created by MakeItTalk, we recommend referring to the original source and related research work. MakeItTalk's creators have conducted thorough evaluations and validations of their generated content, and their findings provide valuable insights into the capabilities and limitations of the framework. Thus, readers interested in a detailed examination of the video output and the intricacies of the MakeItTalk-generated content are encouraged to explore the relevant sections of the MakeItTalk research literature, where a wealth of pertinent information can be found.

Furthermore, it's crucial to highlight the flexibility inherent in our framework. This adaptability goes beyond being confined to a single talking head generation method, providing users with the freedom to integrate a diverse array of methods into their workflow. While MakeItTalk serves as a fundamental component of our research and has demonstrated its effectiveness, our framework is intentionally designed to accommodate a wide range of state-of-the-art talking head generation approaches.

This versatility translates into expanded opportunities for researchers and practitioners in the field. They are not limited solely to using MakeItTalk but have the flexibility to explore, experiment with, and incorporate alternative methods that align with their specific research goals and requirements. By detaching our framework from reliance on a single method, we empower the research community to harness the full spectrum of innovations and advancements in talking head generation. This, in turn, fosters a more dynamic and diverse landscape of possibilities in multimedia content creation, human-computer interaction, and other related domains.

## V. SUMMARY OF RESULTS

The complete experimental analysis was performed on a workstation machine equipped with a XEON E5-1650 v4 3.60 GHz processor, 64 GB of RAM, and 2 GEFORCE RTX 2080 graphical processing units each of which has 12 GB of dedicated graphical video memory, memory bandwidth of 616 GB/second, and 4352 cuda cores.

*A. Synthetic Single 2D Child Imaging Facial Data Results*

In the first phase of the experimental analysis, we used distinct boys' and girls' facial data samples which were shortlisted from one of our previous works where we used StyleGAN to tune ChildGAN [4] models for rendering large-scale child synthetic data. Fig. 4 shows some of the child facial samples rendered using the ChildGAN model. Whereas Fig. 5 shows the face localization and 68 facial landmarks detection results on generated synthetic child data using the Dlib library.

*B. Child Speech Synthesizer Results*

The second part includes generating child audio clips using FastPitch architecture. For now, we have written small text sentences which were then used as input feed data for the FastPitch model. Some of these are as below.

1. "It's raining so we will plan some other day".

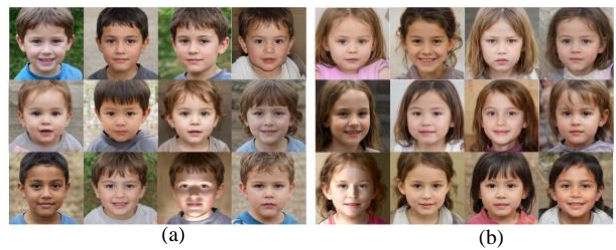

(a)　　　　　　　　(b)

Fig. 4. Generated synthetic child facial subjects, LHS: twelve distinct frontal face samples of boys, RHS: twelve distinct frontal face samples of girls.

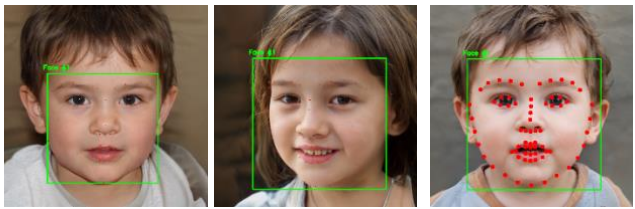

Fig. 5. Face localization and facial landmark detection on ChildGAN data.

2. "Overwhelming majority of people in this country know how to discern and differentiate between what they hear and what they read".

Fig. 6 shows the waveform plot of text ("It's raining so we will plan some other day") to audio generated file using Fastpitch TTS synthesizer text which is generated in the boy's audio voice. The audio is plotted with a 2000 maximum number of sampling points.

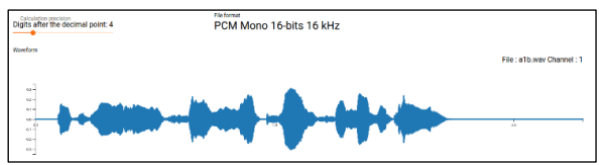

Fig. 6. Audio file plot of the synthesized voice-over of boy generated using Fastpitch TTS.

As mentioned in Section IV-B the generated synthesized voices are further processed by performing a down-sampling operation using the Python librosa library. Fig. 7 demonstrates the graph plot of the original audio WAV file and down-sampled to 16khz audio file.

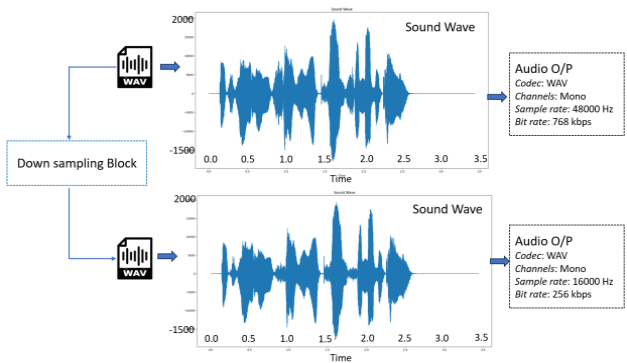

Fig. 7. Downsampled audio WAV file output ("It's raining so we will plan some other day") with a time duration of 3.5 seconds, sample rate of 16000 Hz, and bit rate of 256 kbps.

### C. Single 2D Facial Image to Talking Child Results

The last phase of the experimental results demonstrates the speaking children videos which are rendered using a single 2D RGB frame and driving audio input. Since this work is in the initial phases now, we have rendered outputs of 20 different child subjects. The synthetic speaking children results along with tuned ChildGAN models used are available on our GitHub repository: https://github.com/MAli-Farooq/Synthetic-3D-Speaking-Children. Fig. 8 shows the selected frame-by-frame results extracted from a rendered speaking child video.

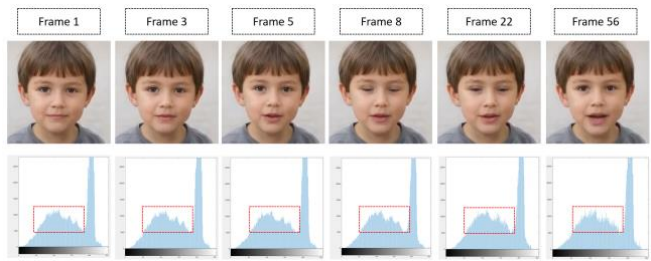

Fig. 8. Five different speaking child video frames of a single subject with their respective histograms.

The second row of Fig. 8 shows the graphical representation of pixel color distribution present in each digital frame which is different from other frames due to continuous lip and eye movement action. Fig. 9 shows the facial landmarks along with varying facial angles of six different facial frames extracted from taking face video results of a similar subject.

### D. Subjective Evaluation of Rendered 3D Child Video Data

The quality of rendered data was further evaluated using human subjective evaluation. For this purpose, we have taken the opinions of six participants from our research group by asking them the following questions.

1. Do you agree that the visual quality of the rendered synthetic child video is good?
2. Do you agree that the audio in the video, including speaker similarity, prosody, and audio quality is good?
3. Do you agree the overall video is of natural quality and sharp?

Among this five participants provided the positive response in favor of $1^{st}$ and $3^{rd}$ questions whereas four participants agreed with question 2. Thus, on average we got a 75% percent positive response ratio from human evaluation.

## VI. CONCLUSION AND FUTURE WORK

Our goal in this work has been to demonstrate the potential of synthetic data for replacing "real-world" data in the particular context of a smart-toy platform. To this end, we have combined several advanced data synthesis techniques to provide a working pipeline for speech-driven animated facial training data samples. While this work is still in its early stages the resulting data samples are convincing, capturing realistic facial features such as synchronized lip and jaw movements and eye blinking. Future work will include quantitative evaluation of the uniqueness of the seed facial data samples, improvements in the quality and number of individual speaker embeddings, and improved controllability of the speech generator (e.g. emotional speech embeddings), and further improvements in the quality and controllability of the facial animations. We are also exploring the addition of a diffusion model to the generator for seed facial samples that will allow specific ethnicities, hairstyles, and facial characteristics to be generated.

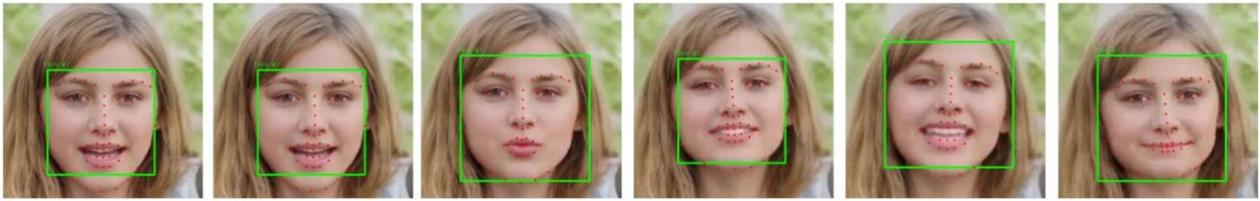

Fig.9: Video frames outputs of girl subject showing varying head pose and lip movements.

ACKNOWLEDGMENTS

This work was supported in part by the DAVID project of the Disruptive Technologies Innovation Fund (managed by the Department of Enterprise, Trade and Employment and administered by Enterprise Ireland), also by the Science Foundation Ireland Centre for Research Training in Digitally Enhanced Reality (www.d-real.ie) under Grant No. 18/CRT/6224, and the ADAPT Centre (Grant 13/RC/2106), and finally by the Irish Research Council Enterprise Partnership Ph.D. Scheme under Grant EPSPG/2020/40.